\documentstyle[12pt,overcite,psfig]{article}

\def\apj{{\it Astrophys.\ J. }}
\def\aj{{\it Astron.\ J. }}
\def\apjs{{\it Astrophys.\ J.\ S. }}
\def\mnras{{\it Mon.\ Not.\ R.\ Astron.\ Soc. }}

\def\araa{{\it Annu. Rev. Astron. Astrophys. }}
\def\rmp{{\it Rev. Mod. Phys. }}
\def\etal{{\it et\thinspace al.}}

\def\hmpc{\,{\rm h^{-1}Mpc}}

\def\kmsmpc{\,{\rm km\,s^{-1}Mpc^{-1}}}
\def\3hmpc{\, ( h^{-1} {\rm Mpc})^3}

\def\omm{\Omega_{\rm m}}
\def\oml{\Omega_\Lambda}

\def\today{\ifcase\month\or
  January\or February\or March\or April\or May\or June\or
  July\or August\or September\or October\or November\or
  December\fi
  \space\number\day, \number\year}

\def\capt{\small \baselineskip 12pt }

\begin{document}

\title{Evidence for a positive cosmological constant\\
         from flows of galaxies and distant supernovae}  

\author{Idit Zehavi$^{*,\dagger}$ \& Avishai Dekel$^*$}
\date{}
\maketitle

\vspace{-30ex}
{\footnotesize \raggedleft FERMILAB-Pub-99/010-A \\}
\vspace{-\baselineskip}
\vspace{30ex}

\vskip 2pt
\noindent $^*$Racah Institute for Physics, The Hebrew University,
     Jerusalem 91904, Israel \\
\noindent $^\dagger$NASA/Fermilab Astrophysics Group, Fermi National 
Accelerator Laboratory, Box 500, Batavia, IL 60510-0500, USA \\
\vskip 2pt

{\bf 
Recent observations\cite{sn_p,sn_r} of high-redshift supernovae seem to
suggest that the global geometry of the Universe may be affected by a
`cosmological constant', which acts to accelerate the expansion rate with
time. But these data by themselves still permit an open universe of low
mass density and no cosmological constant. Here we derive an independent
constraint on the lower bound to the mass density, based on deviations
of galaxy velocities from a smooth universal 
expansion\cite{dek_rees,ND93,ber95,ZZ,FZ}\/.  This constraint rules out
a low-density open universe with a vanishing cosmological constant, and 
together the two favour a nearly flat universe in which the contributions 
from mass density and the cosmological constant are comparable.
This type of universe, however, seems to require a degree of fine tuning
of the initial conditions that is in apparent conflict with `common wisdom'.
}

The standard cosmological model based on Einstein's gravity and 
global homogeneity is characterized by two fundamental 
parameters that measure the main contributions to the total energy density
(in terms of a `critical' density):
the mean mass density $\omm$ and the cosmological constant $\oml$. The  
latter, an intrinsic part of the theory of general relativity, represents 
a uniform energy density that is associated with the vacuum
(not the mass), and, if positive, acts as a repulsive 
component of the global gravitational force field. 
The values of these parameters determine the type of the universe we live in,
its global extent and ultimate fate, as follows.   The parameter plane 
(Fig.~1) is divided by three lines into six regions.
The line $\omm+\oml=1$ defines a `flat' euclidean geometry, 
favoured by theories of inflation in the early universe; it separates 
curved-space (non-euclidean) 
models of `closed' finite extent (above) and `open' infinite 
extent (below).  The horizontal line $\oml=0$ roughly distinguishes between 
an `unbound' universe that will expand forever (above)
and a `bound' universe that will eventually re-collapse (below).
The special point $(\omm,\oml)=(1,0)$ corresponds to 
the simplest model termed Einstein-deSitter; unless the universe 
exactly fits this model, it evolves away from this point as it expands.
If $\oml>0$, the universe eventually crosses 
the line $\oml= 0.5\omm$, which 
separates between a universe whose expansion decelerates 
by attraction (below) or accelerates by repulsion (above). 

New techniques to measure distances of supernovae (type Ia) at large 
distances (cosmological redshifts $0.5 - 1$), where the large-scale 
curvature of space-time plays a noticeable role, enable a classical 
cosmological test based on how the time dependence of the Hubble 
relation between velocity and distance depends on the cosmological 
parameters\cite{carroll}\/.  The corresponding, inclined (blue) confidence 
limits in Fig.~1 are based on the results of the Supernova Cosmology 
Project\cite{sn_p}\/, which are fully consistent with the
findings of the High-z Supernova Search Team\cite{sn_r}\/.
The allowed region in the parameter plane is an elongated stripe, 
crudely approximated by $0.8\omm-0.6\oml=-0.2\pm0.1$ (1$\sigma$ errors). 
Based on this result alone, a nearly flat universe is likely, with 
a broad maximum to the probability near $(\omm,\oml)=(0.3,0.7)$,
but an open model, of $(0.1,0)$, for example, is still allowed at the 
$2\%$ confidence level. An orthogonal constraint is required in order to 
remove the degeneracy between $\oml$ and $\omm$. 

Here we show that such a constraint is provided by 
current data of `peculiar' velocities of galaxies 
on scales of $\sim$300 million light years. These velocities, 
which are produced by gravity due to local fluctuations  
about the mean mass density, depend also on the value of the 
mean density itself, $\omm$. The constraints on $\omm$ from 
these velocities are almost independent 
of $\oml$, but combined with the inclined supernovae constraints 
any bound on $\omm$ effectively serves as a bound on $\oml$.

We use two recent catalogues of galaxy peculiar velocities. One, named 
Mark III\cite{M3}\/, contains distances for $\sim\!3,000$ galaxies within 
$\sim\!70 \hmpc$; the other, named SFI\cite{SFI1,SFI2}\/, consists of 
$\sim\!1,300$ spiral galaxies with a more uniform sampling in a similar volume.
In order to obtain the peculiar velocity of a galaxy, its total velocity 
is measured using the Doppler redshift, and its distance is separately inferred
using the so-called Tully-Fisher method, with an error of $15-21\%$.  
The desired peculiar velocity along the line of sight is obtained by 
subtracting, from the total velocity, the Hubble expansion velocity at the 
galaxy distance. The peculiar velocities are carefully corrected for 
systematic errors\cite{D99,FZ}\/.

\begin{figure}
\vskip -6cm
\hskip -1cm
\centering \mbox{\psfig{figure=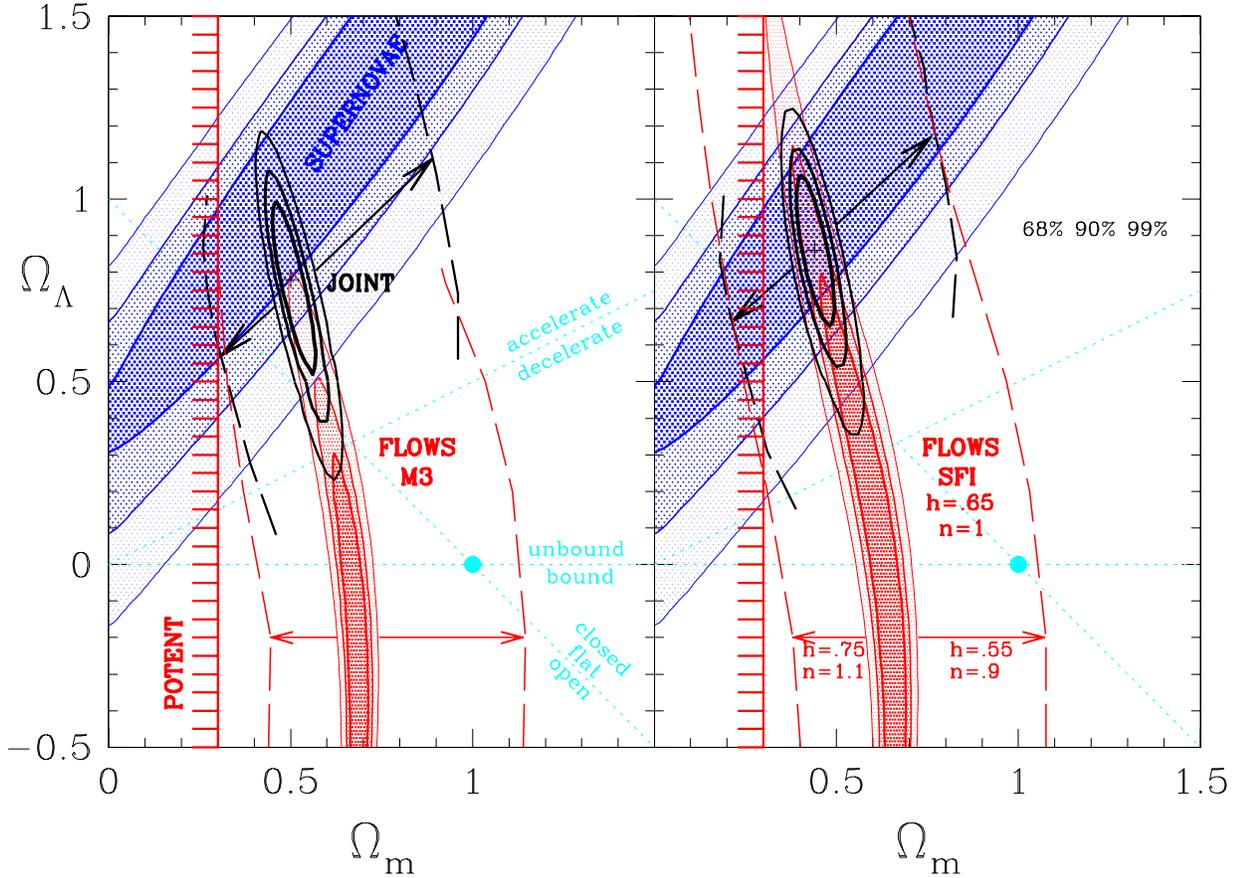,height=23cm}}
\vskip -5cm
\caption[]{\capt
Constraints in the $\omm-\oml$ plane, showing relative confidence limits of 
$68\%$, $90\%$ and $99\%$.  The inclined contours (blue, same in both panels),
that roughly constrain $0.8\omm-0.6\oml$, arise from the global geometry 
of space-time based on supernovae as distance indicators\cite{sn_p}\/.
The lower bound $\omm>0.3$ (red) represents a $99\%$ confidence limit
based on several different studies of peculiar velocities using POTENT 
reconstruction, independent of the biasing relation between galaxies 
and mass\cite{dek_rees,ND93,ber95}\/.
The almost vertical contours (red) arise from the peculiar velocities
of the Mark~III (left) and SFI (right) catalogues, based on a likelihood 
analysis which assumes a parametric mass power spectrum of the CDM family 
with COBE normalization on large scales\cite{ZZ,FZ}\/.
The central contours shown are for a standard case of fixed $h=0.65$ 
and $n=1$; the uncertainties in these values are represented by the 
outer $99\%$ contours on both sides (long dashed red), computed for the 
extreme cases of fixed $(h,n)=(0.55,0.9)$ and $(0.75,1.1)$. These encompass 
a conservative range of non-negligible likelihood based on the flow data. 
The strong bounds from flows are on $\omm$; the apparent upper bound on 
$\oml$ arises indirectly from the COBE normalization and is very uncertain;
the contour can extend further in the $\oml$ direction, allowing significant 
overlap with the supernova confidence region. The corresponding joint 
(relative) confidence limits from supernovae and flows are shown (black 
contours); they are almost the same for Mark III and SFI. 
The combined constraints thus favour an unbound and roughly flat universe 
with comparable contributions from the cosmological constant and mass density.
A universe with very low $\omm$ and zero cosmological constant is ruled out.
}
\label{fig:om_lam}
\end{figure}

The POTENT method\cite{DBF,D99} assumes that the peculiar
velocities were generated by gravity and that they therefore represent a 
potential flow on large scales. This allows a recovery of the 
three-dimensional velocity field of galaxies in our local cosmological 
neighbourhood. For an assumed value of $\omm$, the underlying field of
mass-density fluctuations can be extracted, using slightly nonlinear 
approximations to the relation between velocity and density fluctuations 
within the theory of gravitational instability. Assuming also that the 
initial fluctuations were a gaussian random field, direct dynamical 
constraints on $\omm$ can be obtained without appealing to the observed 
spatial distribution of galaxies. These constraints are thus independent 
of the unknown relation between galaxies and the underlying mass density 
(termed `biasing'). 
Several different methods applied to the POTENT Mark~III fields
consistently yield a lower bound of $\omm > 0.3$ at the $99\%$ 
confidence level, marked by a vertical red line in Fig.~1.
One method constrains $\omm$ based on the large diverging outflow 
recovered in the vicinity of a large nearby underdense region (void), 
using the fact that even an empty void cannot induce outflows as 
large as those observed if the density outside it is too low\cite{dek_rees}\/. 
A second method uses the deviations of the initial fluctuations from a 
gaussian distribution, when derived from the velocity data, assuming 
a value of $\omm$ that is too low\cite{ND93}\/. A third method derives 
$\omm$ based on the gravitationally-induced deviations from gaussian 
distribution of the present-day velocity-divergence field\cite{ber95}\/. 
Similar constraints confirming the lower bound of $\omm > 0.3$ 
are currently being obtained from the SFI data.

In a new analysis\cite{ZZ,FZ}\/, not based on POTENT, we have applied 
a maximum-likelihood analysis to the Mark~III and SFI peculiar velocities 
to determine cosmological parameters. It is done via a parametric model 
for the mass-density fluctuation power spectrum $P_k$, which measures the 
distribution of power among different length scales, proportional to $k^{-1}$.
This analysis uses 
linear gravitational theory and assumes that both the fluctuations and the 
errors are gaussian. The model used is of the popular cold dark matter 
(CDM) model family; $P_k$ is normalized\cite{sugiyama} by the 
fluctuations in the cosmic microwave background (CMB) as measured on very 
large scales by the COBE satellite.  The result provides constraints on 
permissible combinations of three parameters: $\omm$, the power index $n$ 
(where $P_k \propto k^n$ on large scales), and the Hubble expansion 
constant $h$. These parameters enter via the shape of $P_k$ as well as from
the geometry and dynamics of space-time. The constraints obtained from the two 
data sets turn out to be similar; they define a two-dimensional surface 
in the $\omm$-$h$-$n$ space. In the case of a flat cosmology this surface 
can be crudely approximated by $\omm\, h_{65}^{1.3}\, n^2 = 0.58 \pm 0.12$
(where $h_{65}$ is the Hubble constant in units of $65\kmsmpc$, and the 
errors refer to $90\%$ uncertainty).
 
The obtained $P_k$ is determined by the velocity data (on scales 
of $\sim\!50\hmpc$) and is not much
affected by the COBE normalization (at $\sim\!1,000\hmpc$);
a similar $P_k$ is reproduced to within a few percent when the amplitude on 
large scales is left as a free parameter. Also, the result is not dominated 
by the assumed distance errors; $P_k$ is reproduced 
to better than $10\%$ when an error model with free parameters is 
incorporated in the likelihood analysis\cite{FZ}\/. 
This robustness is within the $1\sigma$ limits of the likelihood analysis. 

The peculiar velocities have also been analysed by methods that incorporate 
the spatial distribution of galaxies.  For example, a comparison of the 
Mark III velocities and the IRAS 1.2Jy redshift survey\cite{sigad} 
yields results roughly consistent with the CDM model of $\omm \simeq 0.5$ 
favoured by the $P_k$ analysis of flows reported above. Our current bounds 
on $\omm$ from flows also partly overlap with constraints from the evolution 
of galaxy clusters\cite{eke} ($\omm=0.45\pm0.2$).
However, a comprehensive joint analysis of all the available constraints
is beyond the scope of this work. In particular, a thorough interpretation 
of the results involving redshift surveys 
(which span a noticeable range) must include a full discussion of the 
non-trivial biasing relation between galaxies and mass\cite{DL}\/, 
and of the complex systematic effects involved in these different studies. 
Instead, we choose to focus here on one set of constraints, the 
dynamical constraints from flows, combined with the supernova constraints.

In order to obtain the desired constraints in the $\omm$-$\oml$ 
plane, we have applied the $P_k$ likelihood analysis to the Mark~III and 
SFI velocities, using the COBE-normalized\cite{cobe} CDM models
with varying $\omm$ and $\oml$ that now span the whole parameter plane.
The velocities are expected to be insensitive to the value of 
$\oml$\cite{lahav_lam}\/, but a weak $\oml$ dependence occurs as a
result of the COBE normalization imposed.  This is responsible for 
the slight left bending of the red vertical likelihood ridges (Fig.~1), 
and the apparent upper bounds on $\oml$. The $1\sigma$ error in the 
COBE normalization translates to additional uncertainties of $\pm 6\%$ in 
$\omm$ and $\pm 20\%$ in $\oml$, which are on the order of the uncertainties 
displayed by the likelihood contours shown. Thus, the COBE errors would tend 
to stretch the likelihood contours vertically along the $\oml$ axis, 
weakening the apparent upper bounds on $\oml$. 
 
Since the velocity data favour an extended two-dimensional surface 
in the $\omm$-$h$-$n$ space, the likelihood analysis cannot 
determine the three parameters simultaneously; two of them should be 
fixed initially. We adopt as our standard case the simplest, scale-invariant
($n=1$) initial spectrum, and $h=0.65$ (as favoured by nearby supernova 
data\cite{RPK})\/. We crudely estimate a $\pm 15\%$ uncertainty in the Hubble 
constant\cite{hubble}\/, and $\pm 10\%$ in the power index\cite{bond}\/. 
In order to illustrate the sensitivity of the constraints to these 
parameters, we have also determined $\omm$ and $\oml$ for two extreme cases
of $h$ and $n$ values. The outer $99\%$ contours are shifted accordingly 
in Fig.~1.

The likelihood ridge of SFI extends into higher values of $\oml$ than for
Mark III. The independent constraints from flows (using $P_k$) and supernovae  
overlap significantly for SFI, but only near the $90\%$ contours for Mark III 
with $(h,n)=(0.65,1)$ (which improves when COBE errors are considered or if 
$n>1$ or $h>0.65$).
The large uncertainty in the upper bound on $\oml$ indicates that
the degree of overlap should not be interpreted as indicating either 
consistency or inconsistency between the velocity and supernova data
and the assumed standard cosmological model. 

A joint parameter estimation from supernovae and flows is therefore 
meaningful, although it is limited to  relative likelihoods; a measure of 
absolute probabilities is not straightforward, despite a seemingly acceptable 
goodness of fit. The black joint contours (Fig.~1) are computed by 
multiplying the two likelihood values at each point, having assumed that 
the data are independent. The most likely joint values (shown as crosses
in Fig.\ 1) for the supernovae and 
the combined Mark III and SFI dataset are $(\omm,\oml)=(0.5,0.8)$,
while the conservative joint $99\%$ confidence limits allow $\omm$ values 
in the range $0.3-0.9$ and $\oml$ values in the range $0.1-1.4$.
The constraints obtained separately from Mark III and SFI are very similar.

The constraints from local flows thus remove the degeneracy
in the constraints from the global-geometry test based on supernovae (and
vice versa), and help rule out an open model with zero cosmological 
constant.  A nearly flat universe, with comparable contributions of matter 
and cosmological constant to the total energy density, is likely.
The favoured model is therefore an unbound universe that will accelerate 
forever, although it is impossible yet to determine whether the global 
geometry is flat, open or closed.  

Such comparable contributions from the mass density and the 
cosmological constant represent a puzzling fine tuning, for example, 
because the two parameters, $\omm$ and $\oml$,
are expected to vary with time in opposite senses. The standard 
theory expects the cosmological constant either to vanish or be larger 
by many orders of magnitude\cite{weinberg,carroll}\/.   Perhaps
entropic arguments may be needed to explain such a fine tuning. Although 
the observed constraints are not yet finally confirmed, they already 
seem interesting enough to pose a serious challenge to theoretical physics.

Other constraints in the $\omm$-$\oml$ plane are worth mentioning in
this framework. Constraints consistent with the supernova ridge but of larger 
uncertainty arise from the age of old star clusters\cite{chaboyer} 
versus the Hubble expansion rate.  
Constraints of orthogonal orientation, roughly on $\oml+\omm$, 
can be deduced from the acoustic peaks in the sub-degree angular power 
spectrum of fluctuations in the CMB, as observed from balloons and from the
ground\cite{efst_comp,lineweaver,tegmark_comp,white_comp,bond,lasenby}\/. 
Two CMB satellites planned for the next decade, MAP and Planck, are expected 
to provide more accurate constraints\cite{zaldarriaga,tegmark_comp}\/.  
 
Future peculiar-velocity data are expected to improve the 
accuracy of the constraints. In addition to many new velocities 
based on distance indicators of the Tully-Fisher type, the most promising 
sources of large-scale peculiar velocities in the future will
probably be the siblings of the same supernovae discussed 
above, but at low redshifts; their distances can be measured with $5-10\%$ 
accuracy out to large distances and their sampling density is limited 
only by the patience of the observers.  



\parskip=3ex

\noindent {\bf Acknowledgements.}
We thank the Mark~III and SFI teams, including
D. Burstein, S. Courteau, L.N. daCosta, A. Eldar, S.M. Faber,
W. Freudling, R. Giovanelli, M.P. Haynes, T. Kolatt, J.J. Salzer,
M.A. Strauss, G. Wagner, J.A. Willick, A. Yahil, \& S. Zaroubi.
We thank G.R. Blumenthal and O. Lahav for discussions. 
This research was supported by US-Israel Binational Science Foundation
grant 95-00330 and Israel Science Foundation grant 546/98 at the Hebrew
University, and by the DOE and the NASA grant NAG 5-7092 at Fermilab.

\noindent Correspondence and requests for materials should be addressed 
to I.Z. (email: iditz@fnal.gov) and A.D. (email: dekel@astro.huji.ac.il).

\end{document}